\documentclass[runningheads]{llncs}
\usepackage[T1]{fontenc}
\usepackage{graphicx}
\usepackage{hyperref}
\usepackage{color}

\usepackage{subcaption}
\usepackage{float}
\begin{document}
\title{Understanding How Paper Writers Use AI-Generated Captions in Figure Caption Writing}
%
%\titlerunning{Understanding How Paper Writers Use AI-Generated Captions in Figure Caption Writing}
\titlerunning{How Paper Writers Use AI-Generated Captions}
%\titlerunning{AI-Generated Captions in Figure Writing}
\author{Ho Yin (Sam) Ng\inst{1}\orcidID{0000-0001-9316-6146} \and
Ting-Yao Hsu\inst{1}\orcidID{0009-0008-9082-6039} \and
Jiyoo Min\inst{2}\orcidID{0009-0002-2991-6875} \and
Sungchul Kim\inst{3}\orcidID{0000-0003-3580-5290} \and
Ryan A. Rossi\inst{3}\orcidID{0000-0001-9758-0635} \and
Tong Yu\inst{3}\orcidID{0000-0002-5991-2050} \and
Hyunggu Jung\inst{4}\orcidID{0000-0002-2967-4370} \and
Ting-Hao `Kenneth' Huang\inst{1}\orcidID{0000-0001-7021-4627}}
\authorrunning{H. Y. Ng et al.}
\institute{Pennsylvania State University, University Park PA 16802, USA \\
\email{\{hzn5135,txh357,txh710\}@psu.edu} \and
%Springer Heidelberg, Tiergartenstr. 17, 69121 Heidelberg, Germany
University of Seoul, Seoul, Republic of Korea \\
\email{jessy0618@uos.ac.kr} \and
Adobe Research \\
\email{\{sukim,ryrossi,tyu\}@adobe.com} \and
Seoul National University, Seoul, Republic of Korea \\
\email{hyunggu@snu.ac.kr}
}
\maketitle              % typeset the header of the contribution
\begin{abstract}
Figures and their captions play a key role in scientific publications.
However, despite their importance, many captions in published papers are poorly crafted, largely due to a lack of attention by paper authors. 
While prior AI research has explored caption generation, it has mainly focused on reader-centered use cases, where users evaluate generated captions rather than actively integrating them into their writing. 
This paper addresses this gap by investigating how paper authors incorporate AI-generated captions into their writing process through a user study involving 18 participants. 
Each participant rewrote captions for two figures from their own recently published work, using captions generated by state-of-the-art AI models as a resource.
By analyzing video recordings of the writing process through interaction analysis, we observed that participants often began by copying and refining AI-generated captions. 
Paper writers favored longer, detail-rich captions that integrated textual and visual elements but found current AI models less effective for complex figures.
These findings highlight the nuanced and diverse nature of figure caption composition, revealing design opportunities for AI systems to better support the challenges of academic writing.

\end{abstract}

\section{Introduction}
Scientific publications rely on figures and their captions to convey complex information, making captions crucial for enhancing figure interpretability and accessibility.
Research shows that captions improve comprehension of visuals in educational materials~\cite{koutsikou2015effect}, aid students across expertise levels, and enhance accessibility by guiding readers through complex visualizations~\cite{nolan2021captions}.
Given the importance of captions, researchers have explored using AI, particularly Large Language Models (LLMs), to generate captions for scientific figures~\cite{kim2020answering,liew2022using,huang2023summaries,hsu2021scicap}. 
Most prior work focused on \textit{reader}-centered use cases, assuming users would only read and evaluate pre-generated captions without modifying them. 
However, the prevalence of poor captions in published papers---studies have shown that a significant portion of captions in arXiv papers are unhelpful~\cite{huang2023summaries}---stems from paper \textit{writers}, not readers. 
Many paper writers do not pay attention to crafting effective captions for their figures, resulting in many low-quality captions, even in published papers.

\begin{figure}[t]
    \centering
    \includegraphics[width=0.8\textwidth]{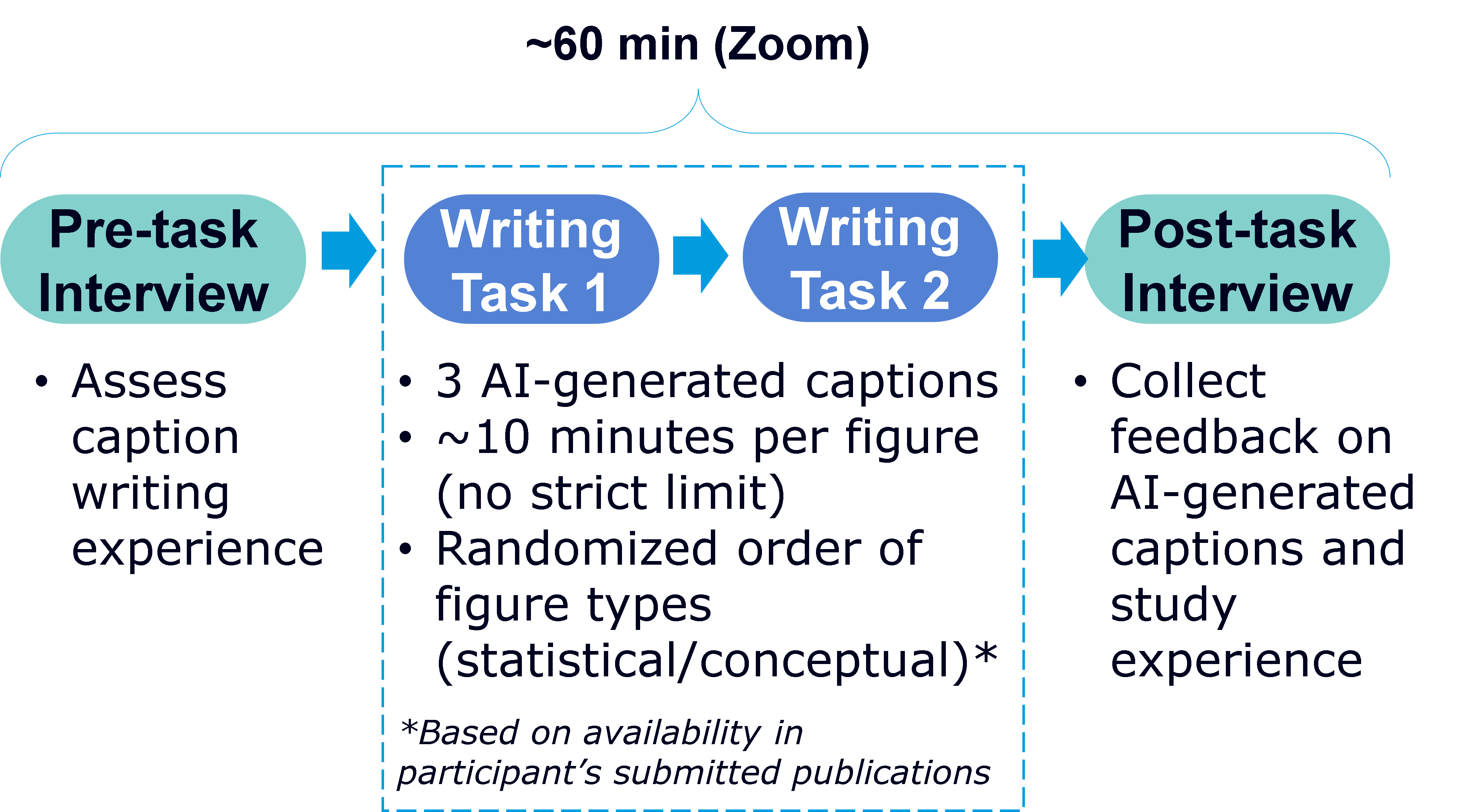}
    \caption{Overview of the user study procedure, which included interviews and two writing tasks. In each task, participants rewrote figure captions from their recently published papers. Participants were provided with the original paper's PDF (with the original caption redacted) and three AI-generated captions to assist in the rewriting process.}
    \label{fig:flowchart}
\end{figure}

This paper examines how \textbf{paper authors integrate captions generated by state-of-the-art figure captioning models into their writing process}. 
Studying the academic writing process presents unique challenges distinct from those involved in supporting academic reading. 
Paper authors are domain experts who have devoted substantial time to their projects, developing a deep understanding of both the content and the broader scientific context. 
This level of expertise often surpasses that of readers of their papers.
This knowledge gap also complicates participant recruitment, as it is challenging to find participants actively writing new papers who are willing to work on their real papers during a user study.
As a result, prior user studies on AI-generated captions often involved participants evaluating captions from papers written by others~\cite{yang2024scicap+,huang2023summaries}.
Even \textsc{SciCapenter}, the only prior work focusing on supporting figure caption writing~\cite{hsu2024scicapenter}, asked participants to write captions for others' papers. 
While this approach simplifies participant recruitment, it does not accurately reflect the real-world scenario where researchers write captions for their own work.
Additionally, scholarly authors face the unique challenge of addressing a broader and more diverse academic audience~\cite{yore2000desired}.
They must consider subtle and nuanced aspects of communication to effectively convey their ideas. 
These considerations are inherently difficult for automated systems to capture and support, making them particularly valuable to study and understand.

This paper presents a user study involving 18 participants, each tasked with rewriting captions for two figures from their own recently published work (Figure~\ref{fig:flowchart}). 
For each figure, participants were provided with captions generated by state-of-the-art AI models, which they could freely use to aid in their caption-writing process. 
During the study, participants had access to the full content of their papers, except that the original captions for the target figures were redacted. 
To study their writing behavior, we recorded the entire caption-writing process on video.
These recordings were then analyzed through interaction analysis~\cite{jordan1995interaction}, which involved manually coding the videos to capture participants' writing behaviors. 
We observed that most participants began by copying substantial portions of the AI-generated captions, subsequently refining or building upon them.
Our findings also suggest that AI-generated captions were more helpful for \textbf{statistical} figures than for \textbf{conceptual}, non-statistical ones.
Additionally, authors showed a preference for longer AI-generated captions that integrated both textual and visual details. 
However, the current AI models struggle to assist effectively with caption writing for complex figures.

Despite being a seemingly narrow aspect of academic writing, figure caption composition involves subtle nuances and exhibits significant diversity across authors. 
These findings highlight opportunities for designing future writing systems that better support the complex and nuanced process of caption writing.

\section{Related Work}
\paragraph{Figure Caption Generation and Evaluation.}
Despite the growing focus on evaluating AI models for generating captions of scientific figures, the evaluation of such models's output often overlooked the writer's perspective. 

Previous studies have widely used automatic metrics to evaluate the quality of AI-generated figure descriptions. 
For instance, BLEU measures word overlap between the generated and reference texts~\cite{10.1145/3341162.3345601,10.1145/3579654.3579667,10.1145/3622896.3622919,10.1145/3302425.3302484,10.1145/3364908.3365287,10.1109/TASLP.2024.3353574,masry-etal-2023-unichart,zhao-etal-2023-investigating,chen-etal-2022-towards-table} and, in a similar vein, ROUGE focuses on how much content from the reference text is included in the generated text~\cite{10.1145/3341162.3345601,10.1145/3399715.3399829,10.1145/3302425.3302484,10.1145/3364908.3365287,zhao-etal-2023-investigating,chen-etal-2022-towards-table}.
METEOR evaluates the generated output by measuring its semantic similarity to reference translations~\cite{10.1145/3341162.3345601,10.1109/TASLP.2024.3353574,chen-etal-2022-towards-table}, considering factors such as synonyms and morphological variations.
However, automatic evaluation metrics have significant limitations.
Human-written captions in scientific papers are often of low quality, leading to unreliable comparisons between machine output and manual captions~\cite{kasai2022transparent}. 
Automatic evaluations do not always align well with human judgments, revealing a gap between machine scoring and actual human comprehension~\cite{deutsch2023ties}. 
Finally, these comparisons miss what was not explicitly stated, including the author's intentions and considerations~\cite{hsu2023gpt}. 

To address these limitations, human evaluations have been incorporated alongside automatic metrics. However, these human evaluations often fail to adequately capture the writer's intent behind the captions~\cite{kantharaj-etal-2022-chart,10.1162/coli_a_00363,liu-etal-2024-chartthinker,obeid-hoque-2020-chart,masry-etal-2023-unichart,zhao-etal-2023-investigating,chen-etal-2022-towards-table}. 
Instead, they focus on aspects like matching degree and reasoning correctness~\cite{liu-etal-2024-chartthinker}, informativeness, conciseness, coherence, and fluency~\cite{obeid-hoque-2020-chart} to understand how well the generated text describes the figures or aids reader comprehension.
These methods, while valuable, miss the unique intentions and considerations of the writers throughout the writing process, including brainstorming, planning, and refinement. 
Motivated by this gap, our work engages real caption authors to rewrite their captions using AI-generated captions, allowing us to understand the AI tool's utility. 

\paragraph{AI Writing Assistants.}
AI writing assistants have been explored in various writing contexts, from creative~\cite{singh2023hide} to legal~\cite{weber2024legalwriter} and medical practice~\cite{escartin2017machine}, highlighting the potential for Human-AI collaboration in specialized tasks. 
In academic writing, AI has been utilized for sentence-level suggestions~\cite{gero2022sparks}, literature review writing~\cite{choe2024supporting,aydin2022openai}, and support in critical evaluation during peer reviews~\cite{sun2024metawriter}. 
The integration of AI writing assistants in academic tasks raises questions about accuracy and trust, with previous studies highlighting both potential benefits and challenges~\cite{buruk2023academic}. 
While many studies have explored AI in academic writing, specific tasks like caption generation remain understudied, particularly from the writer's perspective. 
Our work aims to address this gap by examining how researchers interact with and perceive AI-generated captions in scientific writing.

\paragraph{Human-AI Collaboration.}
Human-AI collaboration has gained significant attention, particularly in academic contexts where AI tools assist in various research and writing tasks.
For example, Shen et al.~\cite{shen2024towards} introduced the concept of bidirectional human-AI alignment, focusing on aligning AI systems with human values while also fostering the adaptation of human practices to AI capabilities.
Lee et al.~\cite{lee2024design} also emphasized the importance of considering both AI capabilities and user writing processes, providing a foundation for our investigation into researchers' perspectives on AI-assisted caption writing for scientific publications.
These works underscore the need for writing tools that not only generate content but also support the user's cognitive processes and decision-making.

\section{User Study}
We conducted a study where 18 participants rewrote captions from their published work using AI-generated captions.
This section provides a detailed overview of the study's design and procedure.

\subsection{Participants}
Participants were recruited through the online questionnaire assessing their academic writing experience. 
The questionnaire collected data on research areas, years of experience in academic writing, English language proficiency, and publication history.
To ensure the quality and relevance of our study, our inclusion criteria targeted researchers with experience in scientific caption writing and at least one recent publication (Details in the `Study Material Preparation' section), including preprints.
Participants were required to upload up to three of their most recent published papers through the online questionnaire. 
We also set a constraint that participants' recent publications should collectively contain at least two figures. We did not impose restrictions on author order to preserve participant anonymity.

We recruited 18 participants representing diverse research areas, with the majority in Computer Science/Informatics~(28\%) and Human-Computer Interaction~(22\%), other also includes Artificial Intelligence/Robotics~(17\%), Energy and Minerals Engineering~(6\%), Mechanical Engineering~(6\%), Environmental Engineering~(6\%), Chemistry/Biochemistry~(6\%), Materials Science~(6\%), Cybersecurity~(6\%). Participants' age ranged from 22 to 44, with 78\% between 26-29 years old.
72\%~ of participants reported that English is not their first language.
All participants had published at least one paper (including preprints) within the past three years.

\begin{figure}[t]
    \centering
    \includegraphics[width=0.8\textwidth]{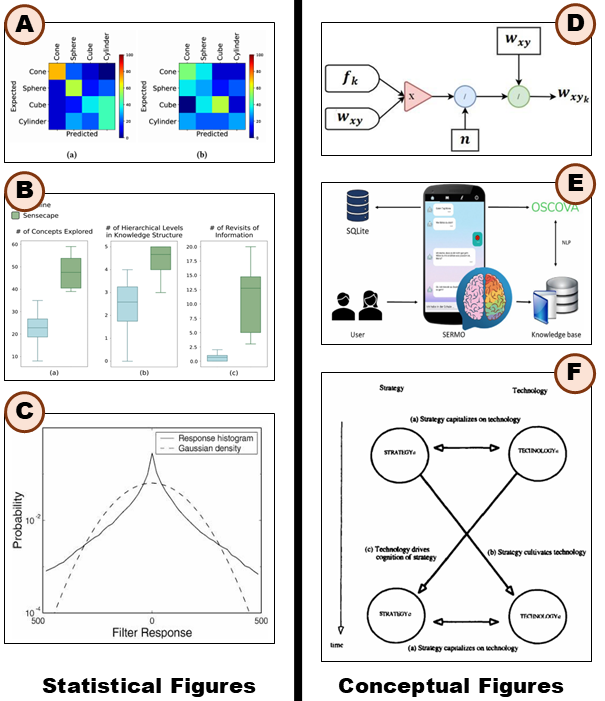}
    \caption{Representative examples of statistical and conceptual figures, with their sources detailed in the Appendix.}    
    \label{fig:exampleFig}
\end{figure}

\begin{figure}[t]
    \centering
    \includegraphics[width=0.95\textwidth]{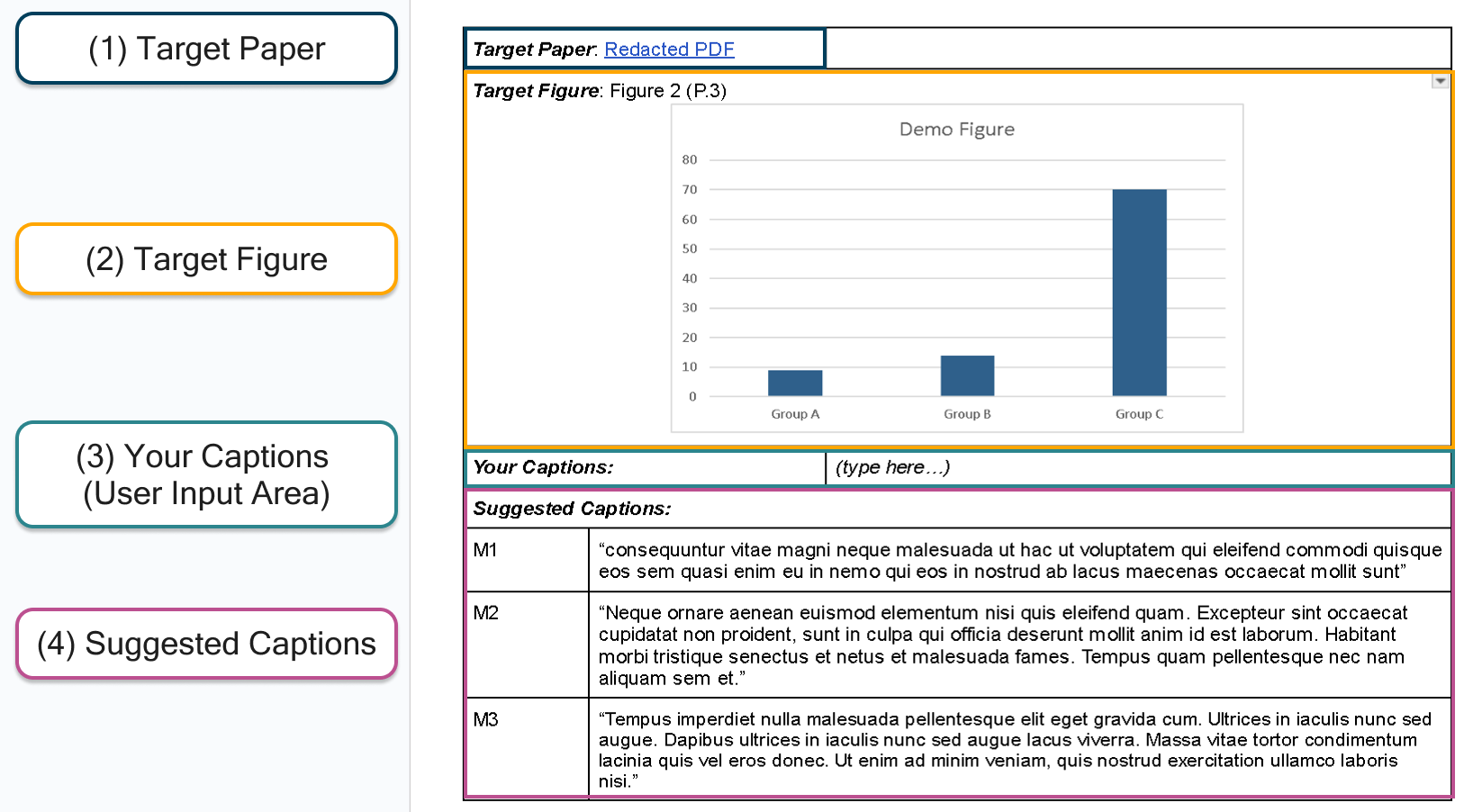}
    \caption{User interface for the figure caption writing task, showing: (1) `Target Paper' - the hyperlink to the redacted PDF, (2) `Target Figure' - displaying the figure image for the writing task, with the figure number and the page number in the redacted PDF, (3) `Your Captions' - User input area for caption writing, and (4) `Suggested Captions' - AI-generated captions from 3 different configurations (\textsc{Unlimited}, \textsc{Text-Only}, \textit{30-Word}), presented in randomized order for each caption item.}
    \label{fig:WritingTaskSetup}
\end{figure}

\subsection{Study Material Preparation}
Participants provided up to three of their own papers published within the last three years, but at least one month ago. 
This timeframe ensured participants were familiar with their work while simulating realistic writing scenarios.

\paragraph{Selecting Statistical and Conceptual Figures from Participants' Papers.}
The selection of figure types aimed to encompass diverse caption writing requirements. 
While prior works have established detailed classifications spanning tables, photos, diagrams, maps, and plots~\cite{dhote2023survey}, 
we simplified the figure classification into two broader categories based on previous work about qualitative and quantitative representation~\cite{borgstede2021quantitative}: 
\textbf{statistical} 
and \textbf{conceptual} figures. 
\textbf{Statistical} figures present quantitative data through graphs or charts, 
while \textbf{conceptual} figures illustrate theoretical models or processes. 
Figure~\ref{fig:exampleFig} shows representative examples.
This binary classification offered a systematic framework for comparing caption-writing strategies across different types of visual information. 
For the study, we aimed to select one statistical and one conceptual figure from each participant's provided papers. 
However, the selection was constrained by participants' research fields, as some areas or venues predominantly feature statistical figures. 
While achieving an equal distribution of figure types across all participants was challenging, we made every effort to balance them.
We ended up including 16 statistical figures (from 15 participants) and 18 conceptual figures (from 16 participants). 
Two figures from one participant were excluded due to their complexity, as they combined both statistical and conceptual elements;
we still completed the user study with this participant and discussed their results in a separate subsection.

\paragraph{Generating Captions for Selected Figures.}
The goal of our study is to understand how paper writers interact with AI-generated captions when composing captions for scientific figures.
For this purpose, we employed state-of-the-art approaches to generate captions for the selected figures in participants' papers.
Prior work has pointed out that scientific figure caption generation is not a straightforward image-to-text task but a generation task that relies heavily on the intertwined scientific context~\cite{huang2023summaries,yang2024scicap+}. 
The context is so crucial that the figure caption generation task can be tackled as a pure text summarization task that ``summarizes'' all the figure-mentioning paragraphs in a paper---without even considering the figure image---into a caption~\cite{huang2023summaries}.
In this study,  we selected GPT-4o, the most recent large vision-language model, as our base model and prompted the model to take both the target figure image and figure-mentioning paragraphs as input to generate figure captions. 
We used a prompt similar to that of SciCapenter~\cite{hsu2024scicapenter}; approaches have been shown to be effective in the SciCap Challenge 2024~\cite{Lee2024,Sun2024,Li2024}.
We prompted GPT-4o in a zero-shot manner (no example provided) to generate captions under three distinct conditions:

\begin{enumerate}

\item \textbf{\textsc{Unlimited}:}
In this setting, we prompted GPT-4o to generate captions using both the figure image and the paragraphs referencing the target figure (e.g., ``Figure 3 shows...'') as input. 
No length constraints were imposed, allowing GPT-4o to produce captions of any desired length.

\item \textbf{\textsc{30-Word}:}
This setting is identical to \textsc{Unlimited}, except that a length constraint of no more than 30 words (tokens) was applied. 
This specific length constraint has also been used in prior work on figure caption generation~\cite{huang2023summaries}.

\item \textbf{\textsc{Text-Only}:}
In this setting, we prompted GPT-4o to generate captions using only the paragraphs referencing the target figure as input, without using the figure image. 
Previous research has demonstrated that figure captions can be generated using textual context alone~\cite{huang2023summaries}.
No length constraints were applied in this setting.

\end{enumerate}

To determine which captions were most preferred, all AI-generated captions were presented to participants in a randomized order, without revealing the model that generated each caption.
Figure~\ref{fig:exampleFig} in the Appendix shows examples.

\subsection{Study Procedure and Setups}

Figure~\ref{fig:flowchart} shows the procedure of our user study, which contained semi-structured interviews and two writing tasks.
The study was conducted remotely via Zoom and lasted approximately 60 minutes.
The entire Zoom session, including participants' computer screens, video, and audio, was recorded for further analysis.
The writing task was conducted in a prepared Google Doc, ensuring a consistent writing environment across all participants while allowing for a natural, unrestricted writing flow. 
For each figure, participants were provided with the following to work on the caption writing: 
{\em (i)} three AI-generated captions presented in randomized order, and 
{\em (ii)} the corresponding PDF file with the original caption redacted.
Participants engaged in the writing task using a \textit{think-aloud} protocol~\cite{fonteyn1993description},
in which the participants verbalize their thought processes and decision-making in real time as they perform the task. 
This enabled us to capture participants' cognitive processes directly, providing insight into their reasoning rather than just their final caption output.

While we estimated that each caption would take approximately 10 minutes to complete, no strict time limit was imposed to ensure a comfortable and natural writing environment. 
After completing each caption, participants were asked to rank the three AI-generated captions based on their perceived usefulness. 
They were then asked to compare their rewritten caption to the original (revealed after writing) by responding to the statement, \textit{``My rewritten caption is better than the original caption,''} using a 5-point Likert scale (1 = Strongly Disagree, 5 = Strongly Agree).

\section{Analysis Methods}
To understand participants' behaviors, we followed the practice of \textit{interaction analysis}~\cite{jordan1995interaction}, a systematic method for examining video recording of human interactions. 
Our analysis involved manual video coding of participants' writing behaviors and their engagement with AI-generated captions. 
Using Dovetail video analysis software,\footnote{Dovetail video analysis software: dovetail.com} we reviewed and coded the recordings by marking specific time segments where participants demonstrated distinct writing activities.
Each video segment was analyzed for verbal think-aloud protocols and writing behaviors, with segments typically defined by natural transitions in participants' activities or explicit shifts in their focus.
We describe our video coding scheme in the following subsection.

\begin{table}
\caption{Coding scheme for AI integration activities in AI-assisted caption writing. Examples for each code are provided in the Appendix.}\label{tab:coding-scheme-ongoing}
\begin{tabular}{lllll}
\cline{1-2}
Code    & Description                                                                                                                                                                                                                                                                                                                                                                                                                                                                                                                                                            &  &  &  \\ \cline{1-2}
 \textsc{Copy}    & \begin{tabular}[c]{@{}l@{}}Participant copies text from an AI-generated caption without any \\ modifications, either by directly copying and pasting or by explicitly \\ referencing it while typing.\end{tabular}                                                                                                        &  &  &  \\ \cline{1-2}
 \textsc{Adapt}   & \begin{tabular}[c]{@{}l@{}}Participant significantly modifies an AI-generated caption by rewording, \\ restructuring, or incorporating their perspective while preserving the \\ original meaning. This may involve editing previously copied text or \\ explicitly referencing an AI-generated caption while typing. The \\ adaptation behavior indicates that participants perceive the underlying \\ content of the AI-generated caption as fundamentally accurate, \\ though requiring stylistic or structural refinement to better suit their needs.\end{tabular} &  &  &  \\ \cline{1-2}
\textsc{Delete}   & \begin{tabular}[c]{@{}l@{}}Participant removes portions of text from an AI-generated caption \\ that they previously copied. These deletions may address redundancy, \\ incorrect information, or unnecessary details.\end{tabular}                                                                                           &  &  &  \\ \cline{1-2}
\textsc{Correct}  & \begin{tabular}[c]{@{}l@{}}Participant modified text from an AI-generated caption that they \\ previously copied to address inaccuracies. These corrections are \\ typically informed by the participant’s understanding of the topic \\ and the conventions of the publication venue.\end{tabular}                                                                                                                                                                                                                                                                    &  &  &  \\ \cline{1-2}
\end{tabular}
\end{table}

\subsection{Code Scheme for AI Integration Activities}
Table~\ref{tab:coding-scheme-ongoing} shows the coding scheme---\textbf{\textsc{Copy}, \textsc{Adapt}, \textsc{Delete}, and \textsc{Correct}}---used in our analysis, which was formed as follows.
Following an inductive coding approach~\cite{thomas2003general}, our coding scheme for AI integration activities emerged through iterative analysis of the video data. 
We initially conducted open coding~\cite{williams2019art} to examine how participants engaged with AI-generated captions during their writing tasks, allowing patterns to emerge naturally rather than imposing predetermined categories.
Through this process, we identified that participants' behaviors could be characterized by their level of reliance on and trust in AI-generated captions. 
These behaviors ranged from direct text copying to slight modifications of AI-generated captions, text corrections, and content deletion, which informed the final coding scheme.

\subsection{Video Coding Procedure}
The AI integration activities code is assigned by tracking each participant's sequence of actions in the draft area.
Each subsequent action is then coded chronologically as the participant continues to develop their draft. 
The coding decisions are informed by both the participants' written output and their think-aloud protocol data.
We systematically coded all sequential actions taken by participants during each writing task, creating a timeline of their writing process. 
This analysis provides a comprehensive understanding of how participants interact with and build upon AI-generated captions in scientific writing.

\section{Results}

\begin{figure}[t]
    \centering
    \includegraphics[width=0.8\textwidth]{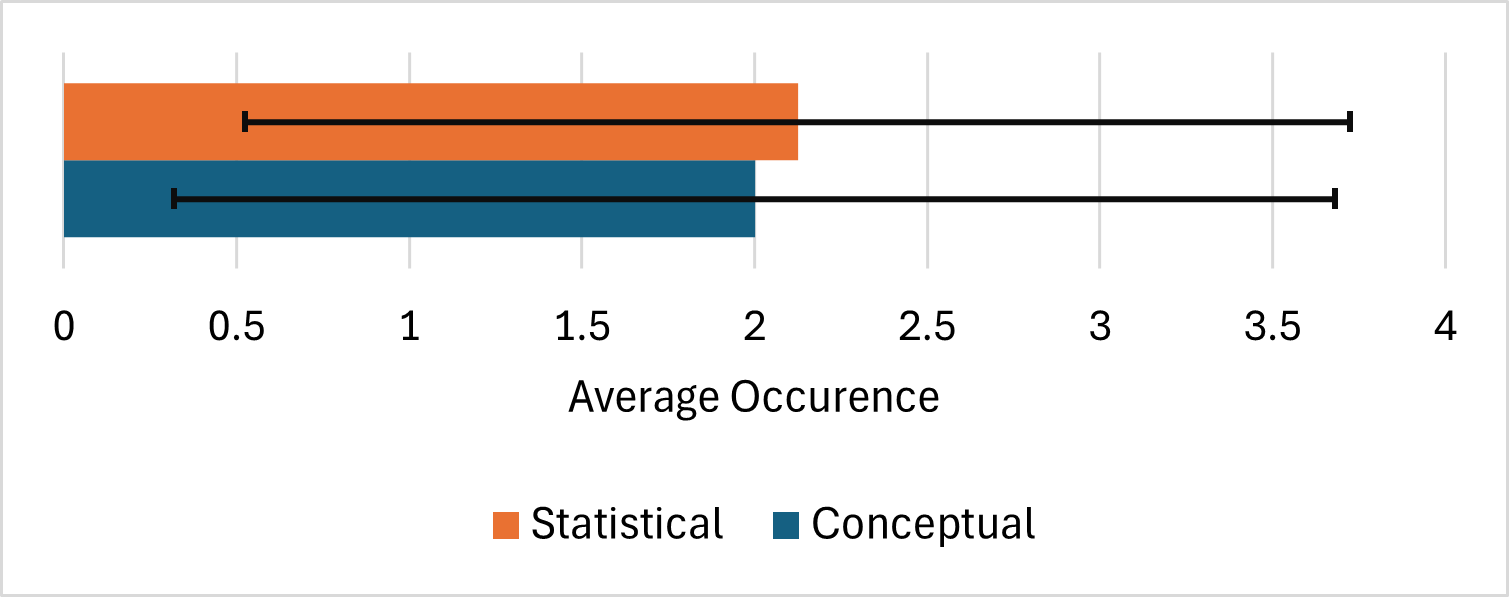}
    \caption{Average number of interactions with AI-generated captions per writing task. Participants interacted with AI-generated captions an average of 2.13 times per session for statistical figures (SD=1.32) and 2.00 times for conceptual figures (SD=1.65).}
    \label{fig:freqByType}
\end{figure}

\begin{figure}[t]
    \centering
    \includegraphics[width=0.8\textwidth]{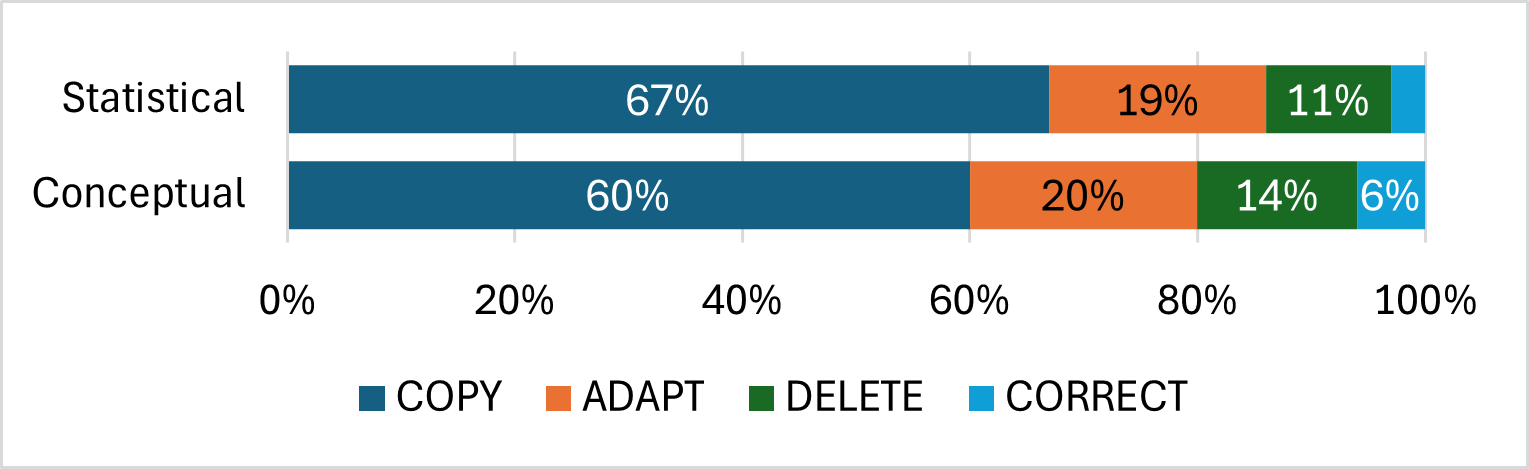}
    \caption{Distribution of the four AI integration activity types for statistical and conceptual figures.}
    \label{fig:adoptionType}
\end{figure}

In our study, we analyzed 16 statistical figures from 15 participants and 18 conceptual figures from 16 participants. 
Two figures from one participant were excluded from most analyses due to their complexity, as they combined both statistical and conceptual elements. 
We still completed the user study with this participant and discussed their results in a separate subsection to explore how AI-generated captions might support caption writing for complex figures.

\subsection{AI-Generated Captions Integration Activities}

\paragraph{Participants rely slightly more on AI-generated captions for statistical figures than for conceptual ones.}
Figure~\ref{fig:freqByType} shows the average frequency of AI integration activities across figure types. 
Participants engaged more frequently with AI generated captions when writing captions for statistical figures (mean=2.13, SD=1.32) compared to conceptual figures (mean=2.00, SD=1.65).
Notably, while participants frequently used AI-generated captions, most instances involved some level of revision.
Across the entire study, only one participant used an AI-generated caption unchanged as the final version for one figure.
Furthermore, we also analyzed the type of integration behavior for each figure type (Figure~\ref{fig:adoptionType}).
\textsc{Copy} was the predominant behavior for both types of figures, with marginally higher frequency in statistical figures. 
Additionally, \textsc{Delete} and \textsc{Correct} occurred more frequently with conceptual figures than statistical figures, suggesting participants were more likely to reject portions of AI-generated captions for conceptual content. 

This difference may reflect LLMs' stronger ability to generate descriptive text versus their limitations in producing abstract or conceptual narratives.
As one participant noted while working on a conceptual figure, ``I'll get rid of this because I don't think we need to emphasize this in this figure''~(P17). 
Another participant emphasized the need for concept revision, stating ``I need to revise the facade because this is not the right way to [present] the dynamic self-design concept.''~(P18)

\begin{figure}[t]
    \centering
    \includegraphics[width=0.8\textwidth]{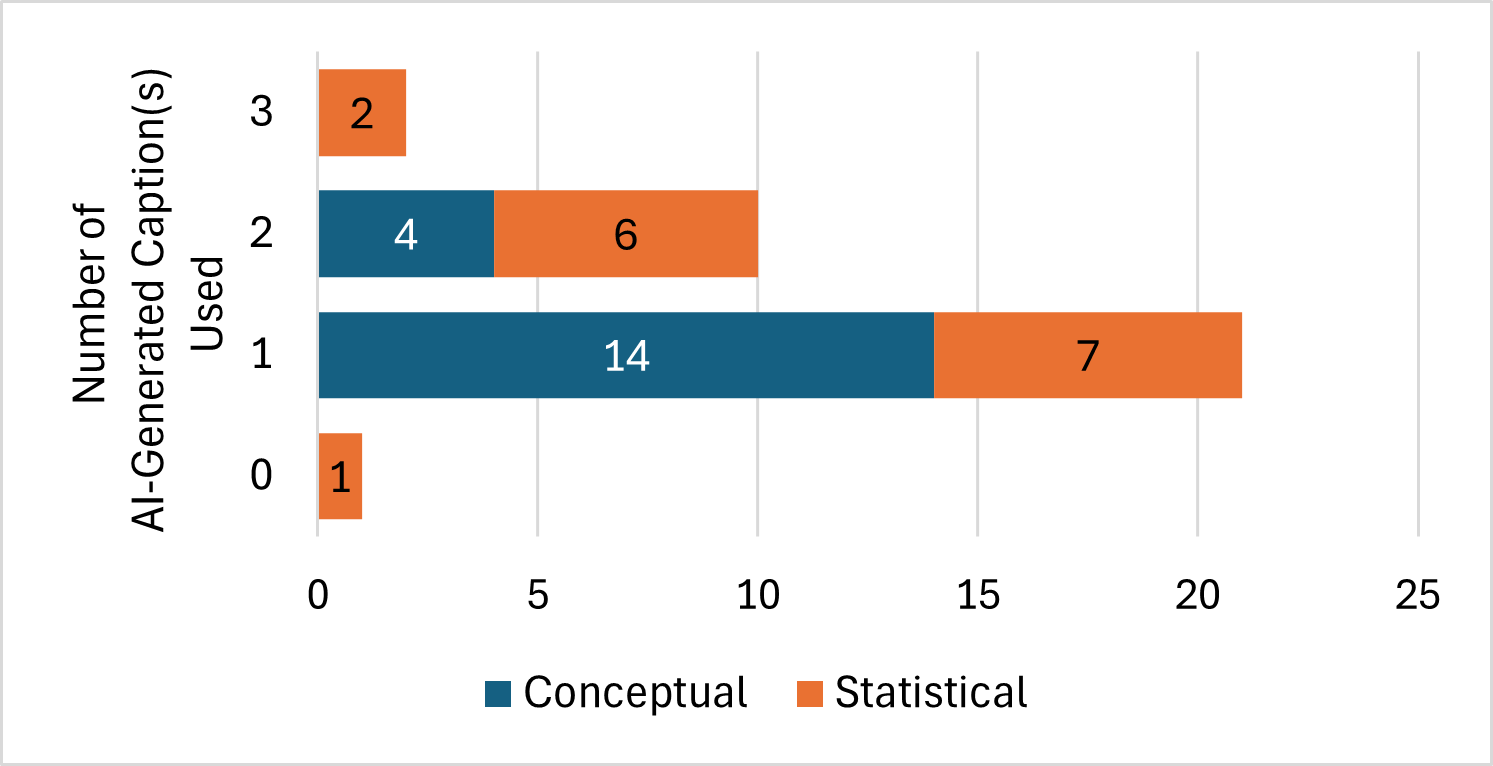}
    \caption{Distribution of the number of AI-generated captions used in a single writing task during the figure caption writing process in our study.}
    \label{fig:SuggestionUsed}
\end{figure}

\paragraph{Most captions were created using a single AI-generated caption, though 35.3\% involved multiple captions.}
Apart from the integration behavior, we also analyzed the number of AI-generated captions used for each writing task.
Figure~\ref{fig:SuggestionUsed} shows the results.
While most captions (21 out of 34) were created using a single AI-generated caption, 12 out of 34 (35.3\%) captions involved multiple AI-generated captions.
Meanwhile, conceptual figures were predominantly completed using one AI-generated caption (14 out of 18), with few cases integrating two AI-generated captions. 
In contrast, for statistical figures, half of the caption items (8 out of 16) were completed using multiple AI-generated captions during the writing process.

\begin{figure}[t]
    \centering
    \includegraphics[width=0.8\textwidth]{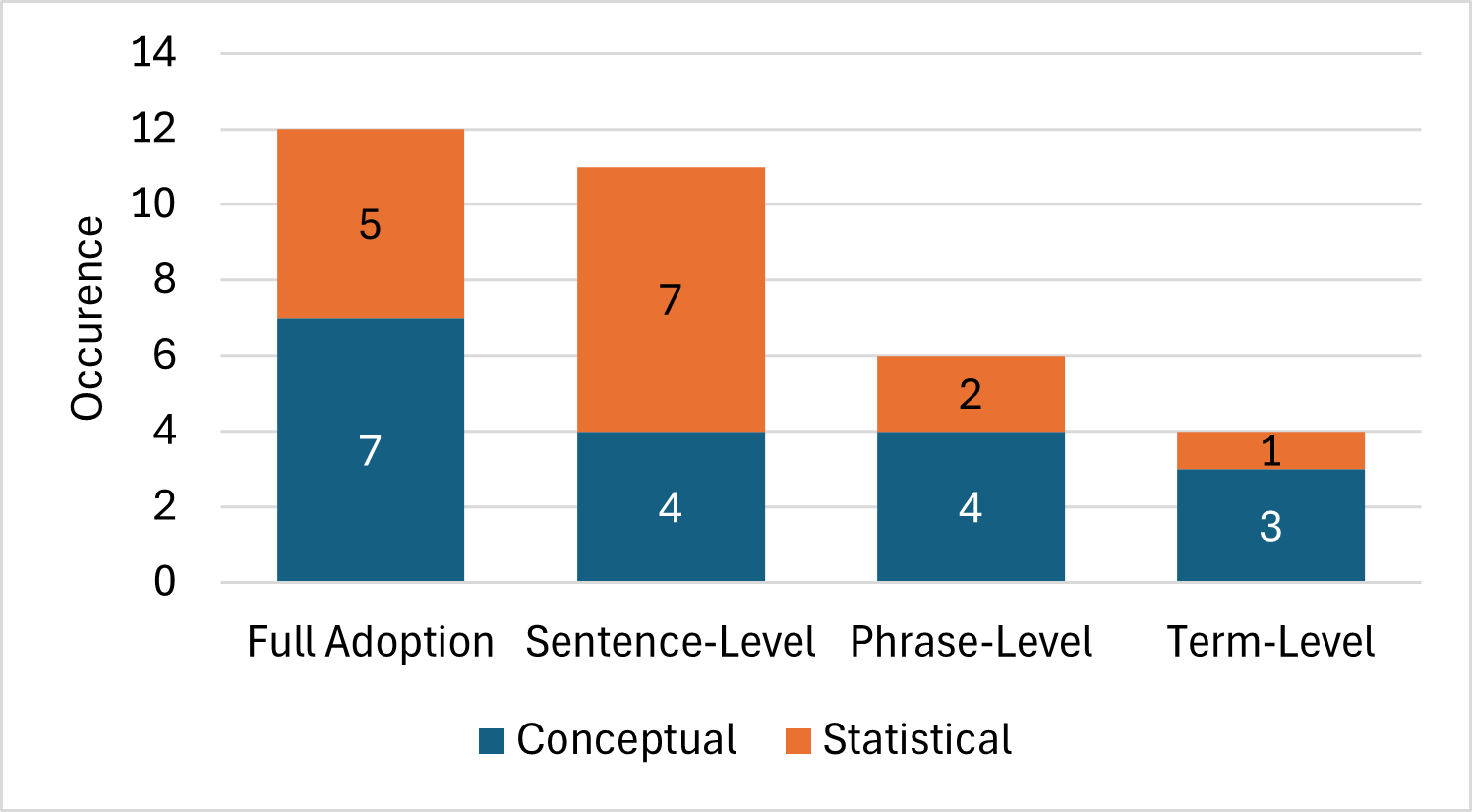}
    \caption{Distribution of AI text usage degrees---\textsc{Full}, \textsc{Sentence}, \textsc{Phrase}, and \textsc{Term}---in participants' initial interactions with AI-generated captions. One session (out of 34) was excluded from this analysis as the participant did not use AI-generated captions.}
    \label{fig:initDist}
\end{figure}

\paragraph{Most participants' primary (first) interaction with AI captions involves using or referencing an entire caption or a full sentence.}
Understanding how much text participants took or referenced from AI-generated captions is crucial but challenging due to the complexity of the writing process, which involves copying, revising, and deleting.
The use of three AI-generated captions further complicates direct comparisons to a single caption.
To simplify the analysis, we focused on participants' \textit{first interaction} with the AI-generated captions. 
On average, participants interacted with the captions twice (see Figure~\ref{fig:freqByType}), and our observations showed that the first interaction often involved the main caption they used in their writing.

This analysis was based on video coding developed using the coding scheme described in Table~\ref{tab:coding-scheme-ongoing}. 
Among the 34 figure-writing sessions, 26 sessions' first interaction with the AI-generated captions was \textsc{Copy}, 7 sessions was \textsc{Adapt}, and only 1 session did not involve any interaction with AI captions.
None of the sessions' first interaction with AI was \textsc{Correct} or \textsc{Delete}.
For each of the 33 \textsc{Copy} and \textsc{Adapt} activities, we further coded the degree of AI text used during the interaction.
The degree of use was categorized into four levels: 
\textsc{Full}, where the entire text of an AI-generated caption was used; 
\textsc{Sentence}, where a complete sentence from the AI-generated caption was used; 
\textsc{Phrase}, where a phrase or text span was taken; and 
\textsc{Term}, where only a single word from the AI-generated caption was used.
Figure~\ref{fig:initDist} shows the results.
Most participants' first interaction with AI captions involved using or referencing an entire caption or a full sentence, which they often revised later to improve its quality. 
Notably, there was only one instance where a participant used an AI-generated caption as-is without any revisions for the final outcome.
The distribution of AI-generated text usage levels was similar for statistical and conceptual figures.

\paragraph{Participants typically start by using AI-generated captions and modify them afterward.}
We conducted a separate analysis of the 34 writing sessions and found that in most cases (31 out of 34), participants began their writing by copying or referencing AI-generated captions before revising them.
In only 3 sessions did participants start by typing without using or referring to any AI content.

\subsection{Model Preferences}
\begin{figure}[t]
    \centering
    \includegraphics[width=0.8\textwidth]{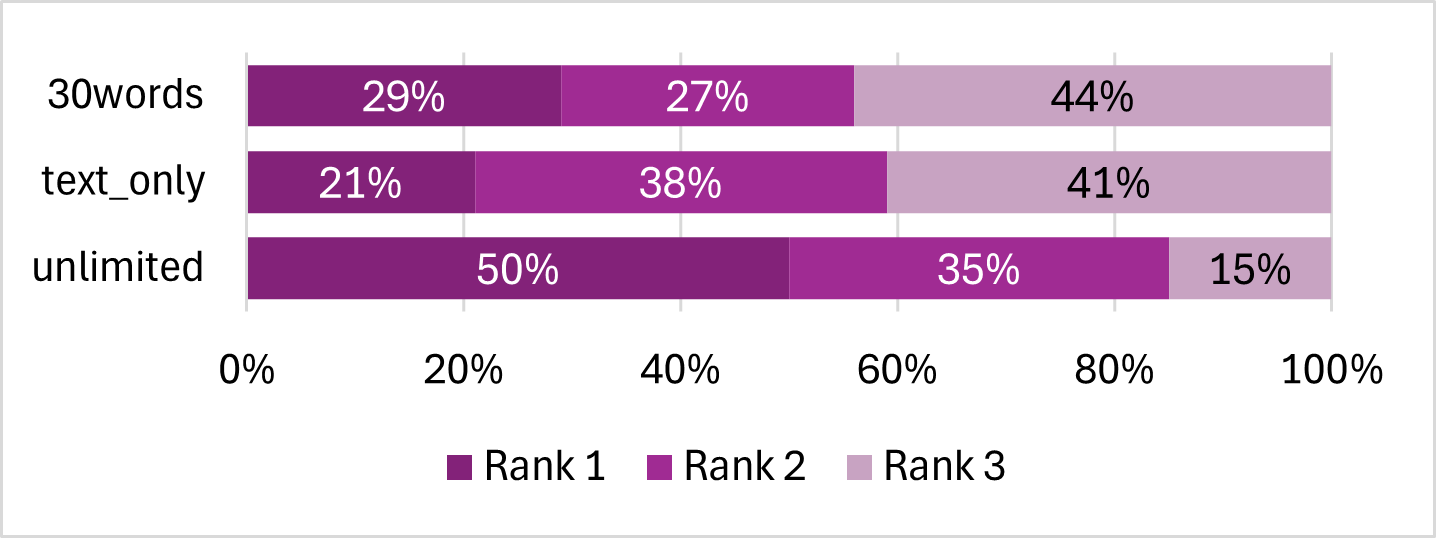}
    \caption{Distribution of GPT-4o configurations ranked by participants, with Rank 1 indicating the best and Rank 3 the worst (N=34).}
    \label{fig:prefModel}
\end{figure}

Participants were asked to rank each AI-generated caption based on its usefulness in helping them compose the final caption.
Figure~\ref{fig:prefModel} summarizes how often captions generated under each of the three conditions were ranked 1st (Best), 2nd, or 3rd (Worst) across all 34 writing sessions.
Captions generated using both figure images and figure-mentioning paragraphs without a length constraint (\textsc{Unlimited}) were ranked first in 50\% of cases.
This suggests that participants may prefer AI-generated captions with more information, as these captions were longer and incorporated both textual and visual details. 
In contrast, captions generated with a 30-word length constraint (\textsc{30-Word}), which also used images and text, were ranked first more often (29\%) than captions generated using only text paragraphs (\textsc{Text-Only}) (21\%).
However, \textsc{30-Word} captions were also ranked worst (third) the most frequently (44\%) among the three conditions.
This indicates that while including both visual and textual information is important, imposing a strict length constraint may significantly reduce the usefulness of generated captions.

\subsection{Perceived Caption Quality Improvement}
At the end of each writing task, participants compared the quality of their rewritten captions to the original captions from their published papers (\textit{``My rewritten caption is better than the original caption''}). 
As a result, 68\% of the captions were rated 4 or 5, with an overall mean rating of 3.82 (SD = 1.24).
However, these results should be interpreted with caution, as the assessment inherently favored the rewritten captions.
Participants had access to both their original work and AI-generated suggestions, enabling them to refine their captions further. 
Additionally, participants were likely more experienced at the time of the study than when they initially wrote the original captions, introducing another potential bias.

%The assessment inherently favored the rewritten captions, as participants had the advantage of both their original work and AI-generated captions to refine their final captions. 
%This bias is amplified by participants being more experienced at the time of the study compared to when they wrote the original captions. 

%Despite this, 
%Our analysis showed that 68\% of caption items received ratings of 4 or 5, with an overall mean rating of 3.82 (SD=1.24).

\subsection{Special Case: Hybrid, Complex Figures}
One participant's published paper included figures that were difficult to classify as either statistical or conceptual, as they contained multiple subfigures combining both statistical and conceptual elements. 
While we completed the user study and analysis for this participant, their results were excluded from the main findings discussed above. 
In this subsection, we examine these two unique cases.

In both writing tasks, the participant relied minimally on AI-generated captions, choosing to write most of the content independently. 
They noted that the AI-generated captions provided limited utility, as the participant noted ``One thing I realized is that it might be described as hallucination because for the first paper [...] the summary was totally different from what I discussed in the main text. So it might not be as accurate''~(P04).
Addressing this challenge presents a critical avenue for future research in developing AI systems capable of handling multi-component academic figures more effectively.

\section{Discussion}
Our results indicate that writers generally preferred captions generated from multimodal inputs with unconstrained lengths, as the additional information provided greater flexibility to choose, add, delete, and modify content. 
This preference for rich information suggests that future AI writing assistants should prioritize interactive suggestions and embrace a human-in-the-loop approach, given that direct adoption of AI-generated content without revision was rare.

\paragraph{Limitations.}
While our paper presents several interesting findings, we acknowledge a few limitations. 
First, our study was not interactive; participants could not iterate with or prompt AI models to revise their captions, which limits the generalizability of our findings. 
Second, participants were asked to rewrite captions for their own published papers. 
While this is closer to real-world caption-writing scenarios than most prior studies---a reasonable trade-off in study design---it still differs from writing captions for new, unpublished work. 
Finally, our generation process relied on the full context of completed and polished papers, whereas in real-world scenarios, AI-generated captions may need to be based on incomplete or even empty drafts.

\section{Conclusion}
This study examined how paper writers use AI-generated captions in scientific caption writing, analyzing the writing processes of 18 participants. 
The results showed that paper writers favored longer, detail-rich captions that integrated textual and visual elements but found current AI models less effective for complex, conceptual figures.
These insights enhance our understanding of human-AI collaboration in academic writing and lay the groundwork for developing AI tools tailored to the unique demands of different figure types. 
Future work should focus on specialized AI systems that adapt to researchers' diverse integration preferences, improving the efficiency and quality of scientific communication while maintaining the human element.

\begin{credits}
\subsubsection{\ackname} We extend our gratitude to the participants of our user study for their valuable contributions and to the anonymous reviewers for their insightful feedback. This research was supported by the Alfred P. Sloan Foundation (Grant Number: 2024-22721).
\end{credits}
 \bibliographystyle{splncs04}
 \bibliography{bib/aaai25}

\newpage
\appendix
\section{Examples of AI-Generated Captions}
\begin{figure}[htbp]
    \centering
    \includegraphics[width=\textwidth]{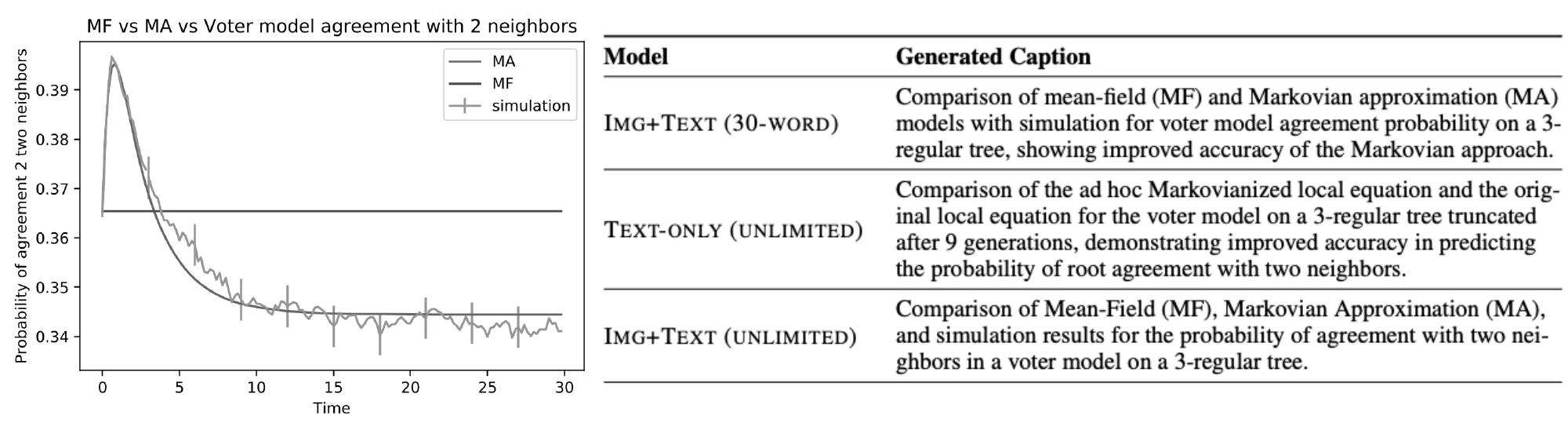}
    \caption{GPT-4o generated captions in three different configurations: (i) GPT-4o (image+text) with a 30-word limit, (ii) GPT-4o (text-only) with unlimited length, and (iii) GPT-4o (image+text) with unlimited length. The figure is sourced from Ramanan~\cite{ramanan2023interacting}.}
    \label{fig:study-sample}
\end{figure}

\section{Code Scheme for AI Integration Activities: Examples.}
\begin{figure}[H]
    \begin{subfigure}{\textwidth}
        \centering
        \includegraphics[width=\textwidth]{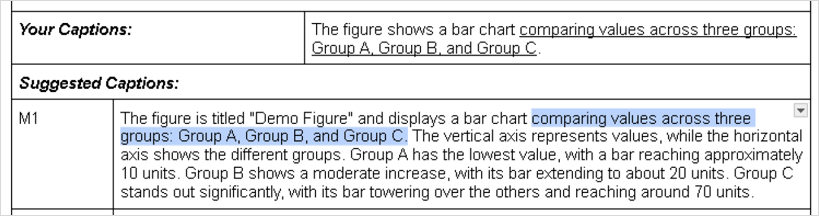}
        \caption{Example of \textsc{Copy} - Retain AI suggestion in its original form}
        \label{fig:sub1}
    \end{subfigure}
    \hfill
    \vspace{1em}
    \begin{subfigure}{\textwidth}
        \centering
        \includegraphics[width=\textwidth]{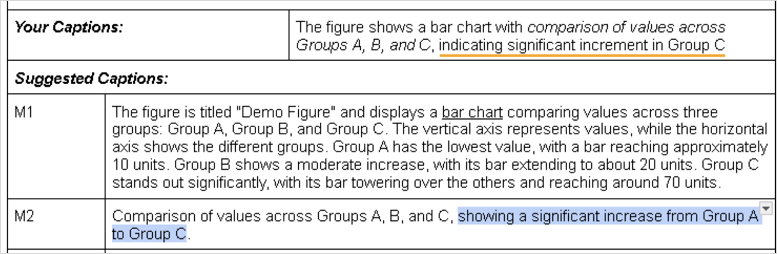}
        \caption{Example of \textsc{Adapt} - Incorporates AI suggestion's core concept while modifying its linguistic expression}
        \label{fig:sub2}
    \end{subfigure}
     \hfill
     \vspace{1em}
    \begin{subfigure}{\textwidth}
        \centering
        \includegraphics[width=\textwidth]{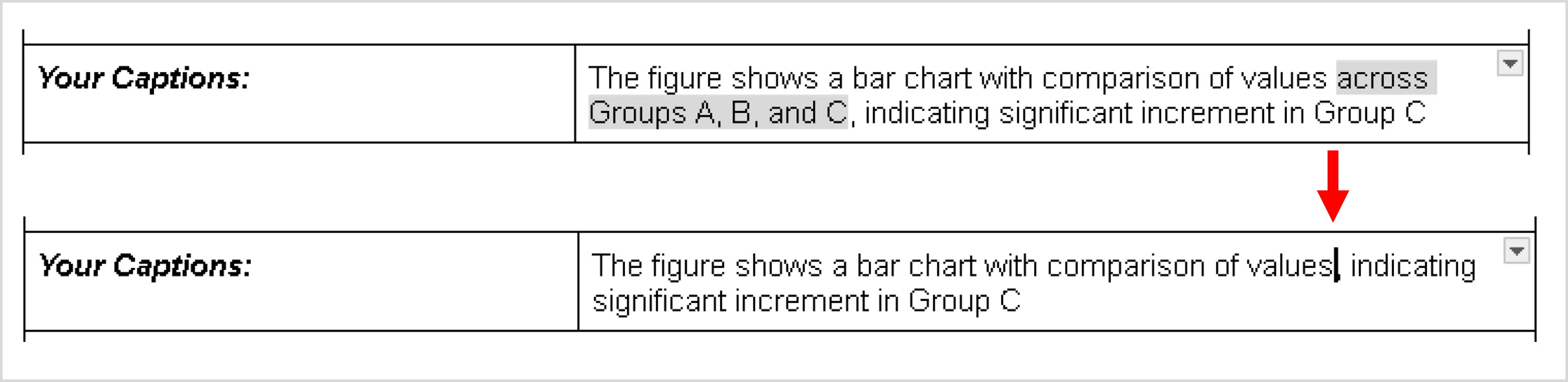}
        \caption{Example of \textsc{Delete} - Discards AI suggestion in draft area}
        \label{fig:sub3}
    \end{subfigure}
    \hfill
     \vspace{1em}
    \begin{subfigure}{\textwidth}
        \centering
        \includegraphics[width=\textwidth]{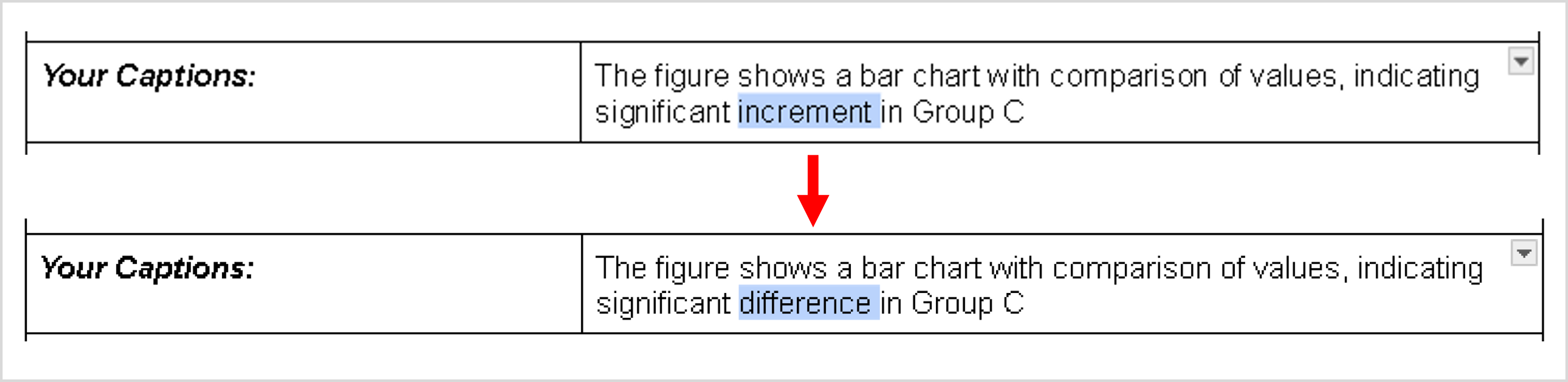}
        \caption{Example of \textsc{Correct} - Modifies AI suggestion to ensure accuracy}
        \label{fig:sub4}
    \end{subfigure}
    \caption{AI Suggestion Integration Activities in Caption Writing: Examples}
    \label{fig:codeExample}
\end{figure}

\section{References for Example Figures in Figure 2}
\begin{itemize}

\item \textbf{Figure~\ref{fig:exampleFig}A:} (Denecke, Vaaheesan, and Arulnathan 2020)~\cite{denecke2020mental}

\item \textbf{Figure~\ref{fig:exampleFig}B:} (Itami and Numagami 1992)~\cite{itami1992dynamic}

\item \textbf{Figure~\ref{fig:exampleFig}C:} (Kreimeier et al. 2019)~\cite{kreimeier2019evaluation}

\item \textbf{Figure~\ref{fig:exampleFig}D:} (Niu et al. 2021)~\cite{niu2021organizational}

\item \textbf{Figure~\ref{fig:exampleFig}E:} (Simoncelli and Olshausen 2001)~\cite{simoncelli2001natural}

\item \textbf{Figure~\ref{fig:exampleFig}F:} (Suh et al. 2023)
~\cite{suh2023sensecape}

\end{itemize}

\end{document}